\documentclass[doublecol]{epl2} 
\usepackage[latin1]{inputenc}
\usepackage{graphicx}
\usepackage{amssymb,amsmath}

\title{Oscillatory instability and routes to chaos in Rayleigh-B\'{e}nard convection: effect of external magnetic field}
\author{Yada Nandukumar \and Pinaki Pal}
\shortauthor{Nandukumar \etal}

\institute{                    
  Department of Mathematics, National Institute of Technology, Durgapur~713209, India\\
  \\
}

\abstract{We investigate oscillatory instability and routes to chaos in Rayleigh-B\'{e}nard convection of electrically conducting fluids in presence of external horizontal magnetic field. Three dimensional direct numerical simulations (DNS) of the governing equations are performed for the investigation. DNS shows that oscillatory instability is inhibited by the magnetic field. The supercritical Rayleigh number for the onset of oscillation is found to scale with the Chandrasekhar number $\mathrm{Q}$ as $\mathrm{Q}^{\alpha}$ in DNS with $\alpha = 1.8$ for low Prandtl numbers ($\mathrm{Pr}$).  Most interestingly, DNS shows $\mathrm{Q}$ dependent routes to chaos for low Prandtl number fluids like mercury ($\mathrm{Pr} = 0.025$).  For low $\mathrm{Q}$, period doubling routes are observed, while, quasiperiodic routes are observed for high $\mathrm{Q}$. 
The bifurcation structure associated with $\mathrm{Q}$ dependent routes to chaos is then understood by constructing a low dimensional model from the DNS data. The model also shows similar scaling laws as DNS. Bifurcation analysis of the low dimensional model shows that origin of different routes are associated with the bifurcation structure near the primary instability. These observations show similarity with the previous results of convection experiments performed with mercury.}

\begin{document}

\maketitle
\section{Introduction}
The study of thermal convection of electrically conducting fluids in presence of magnetic field fascinated researchers for many years due to its relevance in astrophysical and geophysical problems~\cite{chandra:book_1961,proctor:others}. Convection in presence of magnetic field is considered to play a significant role in the formation of sunspots, solar granulation, magnetic field generation of stars and planets. Besides these, magnetoconvection is also important in the engineering applications like crystal growth~\cite{hurle:book}, and nuclear heat exchanger~\cite{gailitis:2002}.  These natural as well as industrial problems involve complex geometry and scientists therefore often consider simplified model of magnetoconvection to understand the basic physics.

Rayleigh-B\'{e}nard convection (RBC)~\cite{chandra:book_1961,bodenschatz:ahlers} is a simplified model of convection and it is studied for many years to understand properties of convection both in presence and absence of magnetic field. The linear theory for Rayleigh-B\'{e}nard magnetoconvection has been extensively developed by Chandrasekhar~\cite{chandra:book_1961}. It is known that the primary instability is greatly effected by the presence of vertical magnetic field~\cite{chandra:book_1961}. On the other hand, horizontal magnetic field does not modify the primary instability and remains same as in the absence of magnetic field~\cite{chandra:book_1961,busse:1983,clever:1989,busse:1989,pesch:2006}. However, secondary as well as other higher order instabilities are significantly affected by the presence of horizontal magnetic field.  

There has been extensive theoretical as well as experimental studies on the effect of external magnetic field on the convective flow of electrically conducting fluids to understand the nonlinear aspects of the problem. Early experiments~\cite{lehnert:nakagawa} considered inhomogeneous as well as homogeneous external magnetic fields and found the stabilizing effects of it on the convective flow. Later a series of theoretical works~\cite{busse:1983,clever:1989,busse:1989,pesch:2006} considered both horizontal as well as vertical external magnetic fields and observed that magnetic field plays an important role in inhibiting the oscillatory convection. 

The effect of horizontal magnetic field on the convective flow of liquid metals has been investigated in several experimental studies~\cite{fauve:1981,fauve:JPL_1984,fauve:1984,burr:2002,yanagisawa:2013}. Fauve et al.~\cite{fauve:1981,fauve:JPL_1984} found that horizontal magnetic field have a stabilizing effect on the convective flow and stronger magnetic field can orient the flow along the magnetic field direction. The phenomena of flow reversal and pattern dynamics in presence of horizontal magnetic field have been experimentally investigated in liquid metals~\cite{yanagisawa:2013}. An interesting two parameter experimental study on the effect of horizontal magnetic field on the routes to chaos in the convective flow of mercury has been carried out by Libchaber et al.~\cite{fauve:1983}. They found that the routes to chaos depend on the magnitude of the external magnetic field as the value of the Rayleigh number is increased. Recently, intermittency routes to chaos has also been reported in a model of magnetoconvection~\cite{macek:2014}.

In this article, we consider RBC of low Prandtl-number fluids in presence of uniform horizontal magnetic field. We perform three dimensional direct numerical simulations of the governing equations under Boussinesq approximation with free-slip boundary conditions and investigate the effect of uniform horizontal magnetic field on the onset of oscillatory instability. Magnetic field is found to push the oscillatory instability onset towards higher Rayleigh number ($\mathrm{Ra}$). The supercritical Rayleigh number for the onset of oscillation shows a scaling with $\mathrm{Q}$ as $\mathrm{Q}^{\alpha}$, $\alpha = 1.8$ for low Prandtl numbers ($\mathrm{Pr}$).  
We then focus our investigation on the effect of magnetic field on the route to chaos in  liquid metals ($\mathrm{Pr}\sim 10^{-2}$). DNS shows a period doubling cascade to chaos for low values of $\mathrm{Q}$ and a quasiperiodic route to chaos for higher values of $\mathrm{Q}$. These results show similarity with the experimental results~\cite{fauve:1983}. To understand detailed  bifurcation structure associated with different routes to chaos, we construct a low dimensional model from the DNS data and perform bifurcation analysis. 

\section{Hydromagnetic system}
We consider a thin horizontal layer of electrically conducting fluid of thickness $d$, kinematic viscosity $\nu$, thermal diffusivity $\kappa$, magnetic diffusivity $\lambda$ and coefficient of volume expansion $\alpha$ kept between two horizontal thermally conducting plates in presence of uniform horizontal external magnetic field $\vec{\bf{B_0}} = (0, B_0, 0)$. The system is heated from below and a uniform temperature gradient $\beta$ between the two plates is maintained. The dimensionless magnetohydrodynamic equations under Boussinesq approximation are given by:  
\begin{eqnarray}
{\partial_t{\bf v}} + ({\bf v}{\cdot}\nabla){\bf v} &=& -\nabla p + \nabla^2 {\bf v} + \mathrm{Ra}\theta\hat{\bf e}_3\nonumber\\
&+& Q\left[\partial_y{\bf b} + \mathrm{Pm}({\bf b}{\cdot}\nabla){\bf b}\right],\label{eq:velocity1}\\
\mathrm{Pr}[{\partial_t \theta}+({\bf v}{\cdot}\nabla)\theta]&=&\left[{\nabla}^2 \theta + v_3\right],\label{eq:theta1}\\
\mathrm{Pm}[{\partial_t{\bf b}} + ({\bf v}{\cdot}\nabla){\bf b} &-& ({\bf b}{\cdot}\nabla){\bf v}] = {\nabla}^2 {\bf b} + \partial_y{\bf v},\label{eq:magnetic1}\\
\nabla{\cdot}{\bf v} &=& \nabla{\cdot}{\bf b}=0,\label{eq:continuity1}
%{\nabla}^2 {\bf b}&=&-\partial_y{\bf v},\label{eq:magnetic field}
\end{eqnarray}
where ${\bf v}(x,y,z,t) \equiv (v_1,v_2,v_3)$ is the velocity field, ${\bf b}(x,y,z,t)=b(b_1,b_2,b_3)$ the induced magnetic field due to convection, $\theta(x,y,z,t)$ the  deviation in the temperature field from the steady conduction profile, $p$ the pressure, $g$ the acceleration due to gravity and $\hat{\bf e}_3$ is the vertically directed unit vector. For nondimensionalization, the units $d$, $d^2/\nu$, ${\nu\beta{d}}/{\kappa}$ and ${B_0\nu}/{\lambda}$ have been used for length, time, temperature and induced magnetic field. Four dimensionless parameters namely $\mathrm{Ra} = ({\alpha \beta g d^4})/({\nu\kappa})$, the Rayleigh number, $\mathrm{Pr} =  {\nu}/{\kappa}$, the Prandtl number, $\mathrm{Pm} =  {\nu}/{\lambda}$, the magnetic Prandtl number and $Q = ({B_0^2d^2})/({\rho_0\nu\lambda})$, the Chandrasekhar number appear in the process, where $\rho_0$ is the reference density of the fluid. The boundaries are assumed to be stress-free, maintained at fixed temperatures and perfectly conducting, which imply,
\begin{eqnarray}
v_3 &=& \partial_{z}v_1 =  \partial_{z}v_2  = \theta = 0 \quad \mbox{and} \nonumber\\
b_3 &=& \partial_{z}b_1 =  \partial_{z}b_2 = 0 \quad \mbox{at}\quad z = 0, 1.\label{eq:bcs}  
\end{eqnarray}
Also, we assumed periodic boundary conditions along horizontal directions. In this paper, we investigate convective flow of low magnetic Prandtl number fluids e.g. liquid metals for which $\mathrm{Pm} \sim 10^{-6}$. Therefore, for simplicity we consider the limit $\mathrm{Pm}\rightarrow 0$~\cite{meneguzzi:1987} and in this limit, the equations~(\ref{eq:velocity1}) and~(\ref{eq:magnetic1}) reduce to:
\begin{eqnarray}
{\partial_t{\bf v}} + ({\bf v}{\cdot}\nabla){\bf v} &=& -\nabla p + \nabla^2 {\bf v} + \mathrm{Ra}\theta\hat{\bf e}_3 + Q\partial_y{\bf b},\label{eq:velocity}\\
{\nabla}^2 {\bf b}&=&-\partial_y{\bf v}.\label{eq:magnetic}
\end{eqnarray}
As a result, the induced magnetic field becomes slaved to velocity field (see equation~(\ref{eq:magnetic})). Now the equations~(\ref{eq:theta1}),~(\ref{eq:continuity1}),~(\ref{eq:velocity}) and~(\ref{eq:magnetic}) together with the boundary conditions~(\ref{eq:bcs}) form the mathematical model of the physical system under consideration. The critical Rayleigh number and wave number for the onset of convection in this case are $\mathrm{Ra_c} = {27\pi^4}/{4}$ and $k_c = {\pi}/{\sqrt{2}}$.  We define another parameter called reduced Rayleigh number by $r = {\mathrm{Ra}}/{\mathrm{Ra_c}}$ for the subsequent sections. 

\section{Direct Numerical Simulations}\label{sec:DNS}
We perform direct numerical simulations of the governing equations~(\ref{eq:theta1}),~(\ref{eq:continuity1}),~(\ref{eq:velocity}) and~(\ref{eq:magnetic}) with boundary conditions~(\ref{eq:bcs}) using the psuedo-spectral code~\cite{verma:Arxiv_2011}.
In the simulation code, vertical velocity ($v_3$), vertical vorticity ($\omega_3$) and temperature ($\theta$) fields are expanded with the set of orthogonal basis functions either with respect to  $\{e^{i(lk_cx+mk_cy)}\sin{(n\pi z)}: l, m, n = 0,1, 2, \dots\}$ or with respect to $\{e^{i(lk_cx+mk_cy)}\cos{(n\pi z)}: l, m, n = 0, 1, 2, \dots\}$ whichever is compatible with the {\it free-slip} boundary conditions, where $k_c$ is the wave number. Therefore, the expanded fields take the following form:
\begin{eqnarray}
v_3 (x,y,z,t) &=& \sum_{l,m,n} W_{lmn}(t)e^{ik_c(lx+my)}\sin{(n\pi z)},\\
\omega_3 (x,y,z,t) &=& \sum_{l,m,n} Z_{lmn}(t)e^{ik_c(lx+my)}\cos{(n\pi z)},\\
\theta (x,y,z,t)&=& \sum_{l,m,n} T_{lmn}(t)e^{ik_c(lx+my)}\sin{(n\pi z)}.
\end{eqnarray}

\begin{figure}[h]
\includegraphics[height=!,width=8.5cm]{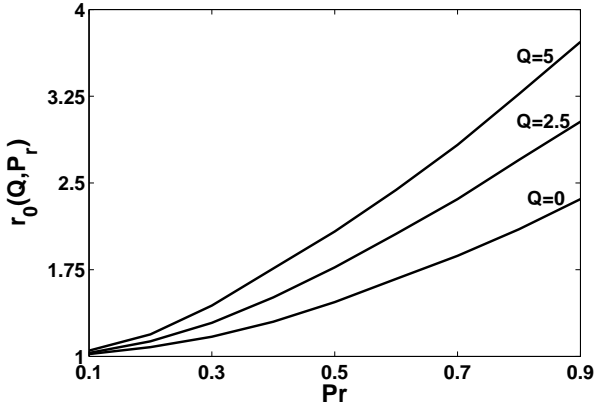}
\caption{The oscillatory instability threshold $r_o(\mathrm{Q},\mathrm{P_r})$ as a function of Prandtl number for three different values of Chandrasekhar number $\mathrm{Q}$ as obtained from DNS.}\label{fig:Pr_r_o(Q0,Q2.5,Q5)}
\end{figure}

The horizontal components of velocity ($v_1$ and $v_2$) are determined from the equation of continuity (\ref{eq:continuity1}) and the induced magnetic field components are then derived from the equation~(\ref{eq:magnetic}). For time advancement, fourth order Runge-Kutta (RK4) scheme is used. The grid resolution for the simulation is taken to be $32^3$ with time step $\Delta t = 0.001$. To test the convergence, some simulations are repeated with $64^3$ grid and found no change in the results.

\begin{figure}[h]
\includegraphics[height=!,width=8.5cm]{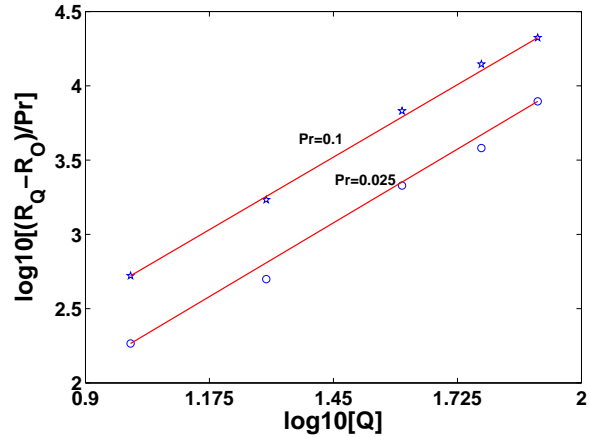}
\caption{Scaling of ${(\mathrm{Ra}_\mathrm{Q}^{o}-\mathrm{Ra}_0^{o})}/{\mathrm{Pr}}$ with $\mathrm{Q}$ for  $\mathrm{Pr} = 0.1$ and $0.025$ as computed from DNS. The solid curves are best fit lines.}\label{fig:Power_law(DNS)}
\end{figure}

\begin{figure}[h]
\includegraphics[height=6.5cm,width=8.5cm]{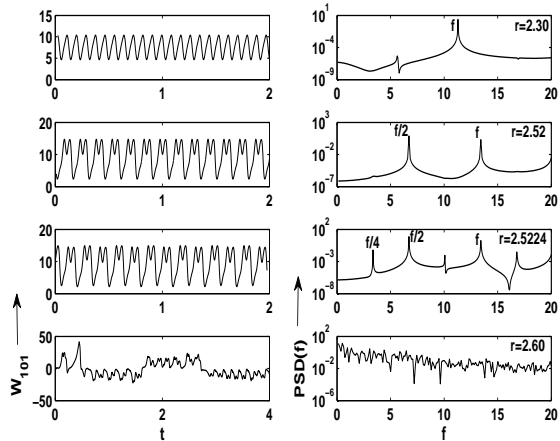}
\caption{Time series as well as PSD of the mode $W_{101}$ computed from the DNS for $\mathrm{Pr}=0.025$ and $\mathrm{Q}=12$.}\label{fig:period2_DNS}
\end{figure}
We first perform DNS of the governing equations to investigate the effect of magnetic field on the onset of oscillatory instability. Fig.~\ref{fig:Pr_r_o(Q0,Q2.5,Q5)} shows the variation of the reduced Rayleigh number threshold $r_0(\mathrm{Q},\mathrm{Pr})$ for the onset of oscillatory instability as a function of Prandtl number for three different values of the Chandrasekhar number $\mathrm{Q}$ as obtained from the DNS. The figure clearly shows that the imposed horizontal magnetic field inhibits the onset of oscillatory instability~\cite{busse:1972}. This result is consistent with the earlier theoretical as well as experimental results~\cite{busse:1989,fauve:1981,fauve:JPL_1984}. 
We compute the relative distance from the threshold of oscillatory instability ${(\mathrm{Ra}_\mathrm{Q}^{o}-\mathrm{Ra}_0^{o})}/{\mathrm{Pr}}$ as a function of $\mathrm{Q}$ from the DNS data for $\mathrm{Pr} = 0.025$ and $0.1$, where  $\mathrm{Ra}_\mathrm{Q}^{o}$ and $\mathrm{Ra}_0^{o}$ are the threshold values of oscillatory solutions in presence $(\mathrm{Q}\neq 0)$ and absence of external magnetic field $(\mathrm{Q}=0)$ respectively. Here we note that for $\mathrm{Pr} = 0.1$, the oscillatory solution bifurcates from steady $2D$ rolls~\cite{busse:1972}, while for $\mathrm{Pr} = 0.025$, it bifurcates from steady cross rolls solution~\cite{mishra:2010}. We find that ${(\mathrm{Ra}_\mathrm{Q}^{o}-\mathrm{Ra}_0^{o})}/{\mathrm{Pr}}$ scales with $\mathrm{Q}$ as $\mathrm{Q}^{\alpha}$, where $\alpha = 1.8$. Fig.~\ref{fig:Power_law(DNS)} represents this scaling as obtained from the DNS for two Prandtl numbers. Similar scaling law was observed in the earlier numerical and experimental investigations~\cite{busse:1983,fauve:JPL_1984}, where they found $\alpha = 1.2$.  The difference in the exponent of the scaling from the earlier investigations may be attributed to the difference in velocity boundary conditions.
 
Now we focus to understand the effect of external magnetic field on routes to chaos of the oscillatory solutions in mercury ($\mathrm{Pr} = 0.025$) as the value of reduced Rayleigh number is increased. In absence of the magnetic field, mercury shows period doubling route to chaos~\cite{fauve:1983}. We find that external magnetic field of smaller magnitude does not change the route to chaos. The time series of $W_{101}$ as well as power spectral density (PSD) plotted in Fig.~\ref{fig:period2_DNS} for four values of $r$ clearly show the period doubling route to chaos for $\mathrm{Q} = 12$ in mercury. On the other hand, DNS for $\mathrm{Q}=150$ show quasiperiodic route to chaos. Now to understand the bifurcation structure associated with $\mathrm{Q}$ dependent routes to chaos in mercury, we construct a low dimensional model from the DNS data.    
\section{Model construction}
From the DNS data we identify $12$ vertical velocity: $W_{101}$, $W_{011}$, $W_{111}$, $W_{202}$, $W_{022}$, $W_{031}$, $W_{301}$, $W_{103}$, $W_{013}$,  $W_{112}$, $W_{211}$, $W_{121}$ , $15$ vertical vorticity: $Z_{100}$, $Z_{010}$, $Z_{110}$, $Z_{111}$, $Z_{112}$, $Z_{310}$, $Z_{130}$, $Z_{120}$, $Z_{210}$, $Z_{102}$, $Z_{012}$, $Z_{201}$, $Z_{021}$, $Z_{121}$, $Z_{211}$ and $13$ temperature: $T_{101}$, $T_{011}$, $T_{112}$, $T_{111}$, $T_{202}$, $T_{022}$, $T_{103}$, $T_{013}$, $T_{301}$, $T_{031}$, $T_{121}$, $T_{211}$, $T_{002}$ most energetic real modes. Now projecting the hydromagnetic equations on these modes, we get a set of $40$ coupled nonlinear ordinary differential equations, which is the low dimensional model for the present investigation.

\section{Analysis of the model and DNS results}
We integrate the low dimensional model using the $ode45$ solver of MATLAB.  First we observe that horizontal magnetic field inhibit the oscillatory instability as observed in the DNS. Then we compute the quantity ${(\mathrm{Ra}_\mathrm{Q}^{o}-\mathrm{Ra}_0^{o})}/{\mathrm{Pr}}$ as a function of $\mathrm{Q}$ for $\mathrm{Pr} = 0.1$ and $0.025$. We find that ${(\mathrm{Ra}_\mathrm{Q}^{o}-\mathrm{Ra}_0^{o})}/{\mathrm{Pr}}$ also scales with $\mathrm{Q}$ as $\mathrm{Q}^{\alpha}$, where $\alpha = 1.3$. 
The exponent found in this case is much closer to the experimental value $\alpha = 1.2$~\cite{fauve:1983}.

\begin{figure}[h]
\includegraphics[height=!,width=8.5cm]{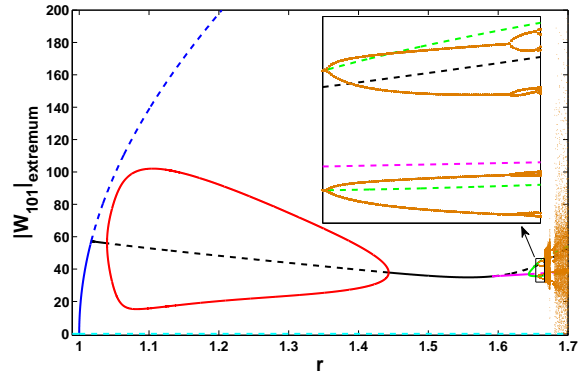}
\caption{The bifurcation diagram computed from the model for $\mathrm{Pr} = 0.025$ and $\mathrm{Q}=20$. The extreme values of $W_{101}$ are plotted as a function of $r$. The solid and dashed cyan, blue, black and pink curves respectively represent stable and unstable conduction state, 2D rolls, CRs I and II. The solid red and green curves represent OCR solutions. The period doubling cascade arising out of the green OCR solutions have been shown with brown dots. Unstbale OCR solutions have been shown with dashed green curve. A zoomed view of the marked region shown in the inset.} \label{fig:bif_Q20} 
\end{figure}

Now we perform the bifurcation analysis of the model, using a continuation software~\cite{dhooge:matcont_2003}. Fig.~\ref{fig:bif_Q20} shows the bifurcation diagram constructed from the model for $\mathrm{Pr = 0.025}$ and $\mathrm{Q} = 20$. In the bifurcation diagram, the extremum values of $W_{101}$ for different solutions have been plotted as a function of $r$ in the range $0.99\le r \le 1.70$. The trivial stable and unstable conduction state have been shown with solid and dashed cyan curve in the bifurcation diagram. The conduction solution or the zero solution is stable for $r < 1$ which becomes unstable via pitchfork bifurcation at $r = 1$. A stable two dimensional ($2D$) rolls solution ($W_{101}\neq 0$ and $W_{011} = 0$ or vice versa) appears. It is shown with solid blue curve in the figure. The stable $2D$ rolls observed in our simulation is due to the choice of the small aspect ratio ($\frac{2\pi}{k_c}: \frac{2\pi}{k_c}$) of the periodicity intervals  as also observed in ~\cite{meneguzzi:1987}. It is interesting to note that Busse and Bolton ~\cite{busse_bolton:1984} had shown that convective rolls with critical wave number  is always unstable for  $\mathrm{Pr} < 0.543$, under the action of long-wavelength instabilities which are  suppressed in our simulation. We also do not observe the secondary instabilities reported in~\cite{busse_bolton:1984}, because of this. 
As the value of $r$ is increased further, the stable $2D$ rolls become unstable via pitchfork bifurcation and stable stationary cross rolls (CR) solution ($W_{101}\neq 0$ and $W_{011}\neq 0$) is generated. The unstable $2D$ rolls are shown with dashed blue curve and stable CR solutions are shown with solid black curve. The appearance of CR from 2D rolls via pitchfork bifurcation  has already been reported by Mishra et al.~\cite{mishra:2010} for $\mathrm{Pr} = 0.02$  in absence of magnetic field and they nicely explained the bifurcation structure associated with the stationary squares (SQ) reported in~\cite{thual:1992}.  For $\mathrm{Pr} = 0.025$, we also observe similar bifurcation structure near the onset of convection.  For low values of $\mathrm{Q}$, the bifurcation scenario is modified but appearance of CR from stationary $2D$ rolls is observed in our simulation.

The CR solutions become unstable via supercritical Hopf bifurcation (HB) at $r = 1.0398$ and stable limit cycles are generated (solid red curves in Fig.~\ref{fig:bif_Q20}). Physically these limit cycles represent oscillatory cross rolls solutions (OCR). The unstable CR solutions continue to exist (dashed black curve) in the regime of stable OCR solutions. It is apparent from the bifurcation diagram that the size of the limit cycles of the OCR solutions initially grows and then decrease with the increase of the value of $r$ and eventually the unstable CR solution becomes stable at $r = 1.4434$, via inverse HB. Again stable CR is observed in the regime $1.4434 \le r \le 1.5925$ (solid black curve). This CR branch again undergoes a pitchfork bifurcation at $r = 1.5925$ and another stable CR branch comes out (solid pink curve). Note that the black CR solution exists as unstable solution. The pink CR solution then undergoes a supercritical HB and stable limit cycle (solid green curve) is generated. This limit cycles undergoes a period doubling cascade (brown dots) as the value of $r$ is increase and eventually chaos is observed. A zoomed view of this period doubling cascade has been shown in the inset. 

\begin{figure}[h]
\includegraphics[height=!,width=8.5cm]{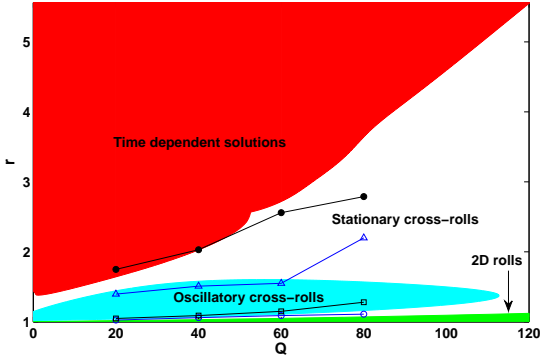}
\caption{Flow regimes on the $\mathrm{Q} - r$ space for $\mathrm{Pr} = 0.025$. The green region represents the stable 2D rolls. CR solutions are represented by white region. An island of OCR has been shown with cyan colour. The red rigion represents time dependent regime. The 2D rolls, OCR, CR and time dependent solutions boundaries obtained from DNS have been dipicted with blue circles, black squares, blue triangles and filled black circles respectively.} \label{fig:phase_diagram}
\end{figure}

We now explore the ($\mathrm{Q}, r$) parameter space for $\mathrm{Pr} = 0.025$ and compute a phase diagram (Fig.~\ref{fig:phase_diagram}) from the model. The diagram delimits different solution regimes of the model with different colours. To describe the phase diagram, first we note that the cyan island exist for $0\leq \mathrm{Q} \leq 112.6$. The regime $0\leq \mathrm{Q} \leq 112.6$ can also be subdivided into two parts namely $0\leq \mathrm{Q} \leq 60$ and $60 < \mathrm{Q} \leq 112.6$. 
For $0 \leq \mathrm{Q} \leq 60$, as the value of $r$ is increased from $r = 1$, a bifurcation structure similar to $\mathrm{Q} = 20$ is observed. This means for $\mathrm{Q} \leq 60$, as the value of $r$ is raised slowly from $r = 1$, stable 2D rolls regime appears via supercritical pitchfork bifurcation of the conduction solution at $r = 1$ (green regime in Fig.~\ref{fig:phase_diagram}). Note that 2D rolls regime becomes very thin as $\mathrm{Q} \rightarrow 0$. As the value of $r$ is increased further, CR regime (white region in Fig.~\ref{fig:phase_diagram}) is stabilized after pitchfork bifurcation of the 2D rolls solutions. Then a pair of supercritical HBs, one forward and the other reverse, occur to the CR solutions. These are associated with the lower and upper boundaries of the cyan island. 
\noindent In between these two HBs, oscillatory cross rolls solutions are observed. After the reverse HB at the upper boundary of the cyan region, the stationary CR regime (white region in  Fig.~\ref{fig:phase_diagram}) again becomes stabilized. For a further increase in the value of $r$, supercritical HB occurs at the lower boundary of the red region in the Fig.~\ref{fig:phase_diagram} and time dependent regime starts. Near the lower boundary of the red region, periodic oscillatory cross rolls solutions are observed. These periodic solutions become chaotic via period doubling route for $\mathrm{Q} \leq 60$ for higher $r$. Note that similar period doubling route to chaos is also observed in DNS for low values of $\mathrm{Q}$. 

For $\mathrm{Q} > 60$, the bifurcation sequence till the value of $r$ reaches the lower boundary of the red region is similar as $\mathrm{Q} \leq 60$. As the value of $r$ is increased further, quasiperiodic transition to chaos occurs. A similar quasiperiodic route to chaos is also observed in DNS for moderate values of $\mathrm{Q}$. The bifurcation sequence observed in DNS is also similar to model. Now we emphasize that DNS also shows qualitatively model like flow regimes and bifurcation structure in ($\mathrm{Q} - r$) parameter space for $\mathrm{Pr} = 0.025$. To show this we compute the bifurcation boundaries for $\mathrm{Q} = 20, 40, 60$ and $80$ and plotted in the Fig.~\ref{fig:phase_diagram}. The blue circles, black square, blue triangle and filled black circles respectively represent branch points, and three HB points obtained from DNS in the Fig.~\ref{fig:phase_diagram}. Note that for $\mathrm{Q} = 100$, we observe only one Hopf for higher value of the Rayleigh number in DNS and the stable limit cycle solution generated out of it becomes chaotic via quasiperiodic route as the value of the Rayleigh number increased further. This implies that the oscillatory cross rolls regime in DNS corresponding to the cyan colored portion in the Fig.~\ref{fig:phase_diagram} terminates for a value of $\mathrm{Q}$ between $80$ and $100$ in DNS. 

The bifurcation scenario for $\mathrm{Q} > 112.6$ becomes qualitatively different in the model. The cyan regime in the Fig.~\ref{fig:phase_diagram} does not exist any more. Two different CR regimes intercepted by cyan regime in Fig.~\ref{fig:phase_diagram} is now connected and only CR is observed in a large range of $r$. The stationary CR solutions becomes unstable via supercritical HB at the lower boundary of the red region and periodic oscillatory solutions are born. These periodic solutions become chaotic via quasiperiodic route as the value of $r$ is increased further. In the model, we observe similar bifurcation scenario till $\mathrm{Q} = 250$ for $\mathrm{Pr} = 0.025$. We have not considered $\mathrm{Q} > 250$ in the model, as beyond this, the model results start deviating significantly from the DNS. 

\begin{figure}[h]
\includegraphics[height=!,width=8.5cm]{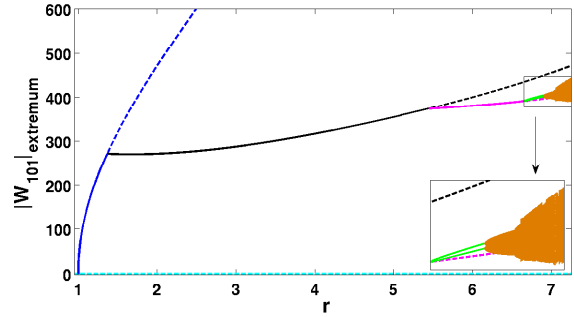}
\caption{Bifurcation diagram computed from the model for $\mathrm{Pr} = 0.025$ and $\mathrm{Q}=250$. The extreme values of $W_{101}$ are plotted as a function of $r$. The stable and unstable stationary 2D rolls, CRs of type I \& II solutions are represented by solid and dashed blue, black and pink curves respectively. Solid green curves represent the OCR solutions. The quasiperiodic solutions are represented by brown dots. A zoomed view of the marked region is shown in the inset.} \label{fig:bif_Q250} 
\end{figure}

To understand the details of the bifurcation structure for the higher values of $\mathrm{Q}$, we construct a bifurcation diagram (Fig.~\ref{fig:bif_Q250}) from the model for $\mathrm{Q} = 250$ with $\mathrm{Pr} = 0.025$. We observe that the bifurcation structure is drastically modified for higher values of $\mathrm{Q}$. The trivial conduction state is stable for $r < 1$ and becomes unstable at $r = 1$ via a pitchfork bifurcation. The stable and unstable conduction solutions are shown with solid and dashed cyan curves in Fig.~\ref{fig:bif_Q250}. Stable 2D rolls (solid blue line in  Fig.~\ref{fig:bif_Q250}) solution is originated. 
\begin{figure}[h]
\includegraphics[height=!,width=8.5cm]{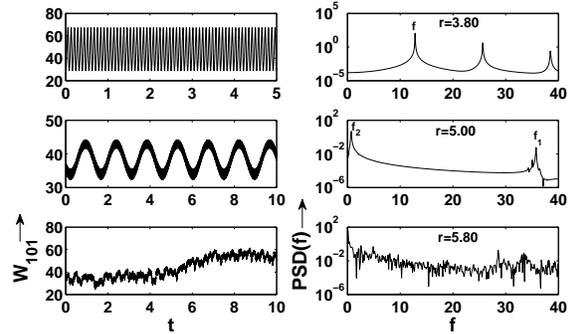}
\caption{Quasiperiodic route to chaos as obtained from DNS for $\mathrm{Q} = 400$ and $\mathrm{Pr} = 0.025$.} \label{fig:time_series_Q400}
\end{figure}
The 2D rolls solution again undergoes a pitchfork bifurcation at $r = 1.368$ and stable CR solution branch is born (solid black curve in Fig.~\ref{fig:bif_Q250}). This CR solutions remains stable for a large range of $r$ and looses stability via HB at $r = 6.652$. Stable limit cycles are generated (solid green curves in Fig.~\ref{fig:bif_Q250}). As $r$ is increased further, the limit cycles become unstable and two frequency quasiperidic solutions are observed (light brown dots in Fig.~\ref{fig:bif_Q250}). The qusiperiodic solution becomes chaotic for higher values of $r$. We observe similar bifurcation sequence for $\mathrm{Q} = 100$, $150$, $200$ $250$, $300$ and $350$ in DNS.
\begin{figure}[h]
\includegraphics[height=!,width=8.5cm]{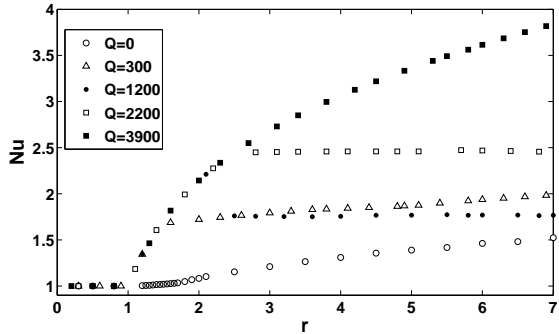}
\caption{Nusselt number (Nu) as a function of $r$ for various $\mathrm{Q}$ with $\mathrm{Pr} = 0.025$ as obtained in DNS.} \label{fig:nusselt}
\end{figure}

We now explore high $\mathrm{Q}$ regime in DNS. DNS for $\mathrm{Q} = 400$ shows quasiperiodic route to chaos but with qualitatively different bifurcations scenario. In this case, the regime of stable $2D$ rolls is greatly enhanced and it does not bifurcate to CR. The $2D$ rolls become unstable via HB and periodic wavy rolls are generated and eventually becomes chaotic via quasiperiodic route for value of $r$. The time series of the mode $W_{101}$ have been shown in Fig.~\ref{fig:time_series_Q400} for three different values of $r$ with $\mathrm{Q} = 400$. The figure clearly shows a quasiperiodic route to chaos. We then consider values of $\mathrm{Q}$ upto $3900$ and found quasiperiodic routes to chaos only in DNS. It is interesting to note that we do not observe subharmonic route to chaos like~\cite{fauve:1984} for high values of $\mathrm{Q}$ in our simulation. The Nusselt number plot for DNS with high $\mathrm{Q}$ have been shown in Fig.~\ref{fig:nusselt}. From the figure, a sharp change in the slope of Nusselt number is observed for $\mathrm{Q}\neq 0$. The heat flux corresponding to the stationary solutions follow a common upper boundary with higher slope and it deviates from the upper boundary as flow becomes time dependent for all nonzero $\mathrm{Q}$.

\section{Conclusions}
In this article, we have presented the results of our investigation on the effect of external horizontal magnetic field on oscillatory instability and routes to chaos in RBC of low Prandtl-number fluids with free-slip boundary conditions. We perform direct numerical simulations as well as low dimensional modeling for the investigation.   We find that oscillatory instability is inhibited due to the horizontal magnetic field and the supercritical Rayleigh number shows a scaling with $\mathrm{Q}$ similar to experimental result. DNS results also show that the routes to chaos depends on the magnitude of external magnetic field. Then we carry out detailed investigation with $\mathrm{Pr} = 0.025$, the Prandtl number of mercury by DNS and low dimensional modeling. A large range of $\mathrm{Q}$ ($0 < \mathrm{Q} \leq 3900$) has been considered for the investigation. We find that the route to chaos is period doubling for lower values of $\mathrm{Q}$ and quasiperiodic routes are observed for higher values of $\mathrm{Q}$. These results show similarity with the experimental results~\cite{fauve:1983}. \\

\end{document}